# A hybrid meta-heuristic approach for channel estimation in OFDM MIMO

Shahriar Hassan[1]*, Umme Farhana[1], Md. Karam Newaz[1]

## Abstract

*In wireless communication Multiple Input Multiple Output (MIMO) technology has brought significant improvement in service by adopting Orthogonal Frequency Division Multiplexing (OFDM), a digital modulation technique. To achieve great performance with MIMO efficiently gathering channel state information (CSI) plays a vital role. Among different approach of channel estimation techniques data-aided channel estimation is more reliable. The existing methods of data-aided channel estimation are Least Square (LS) and Minimum Mean Square Error (MMSE) methods which do not achieve a great performance. Moreover, MMSE is little complex and has higher computational cost. That is why many attempts have been done previously to optimize the methods with help of meta heuristics and also other ways. In this paper we have tried to optimize LS estimation with a combined algorithm of Genetic Algorithm (GA) and Particle Swarm Optimization (PSO). The proposed algorithm has outperformed LS and MMSE. And it gives similar result if we optimize LS with standard PSO but in less numbers of iteration.*

## 1. Introduction

Fast and efficient communication has a greater role in advancement of our lifestyle. As the modern technology advances the need of high data rate in wireless communication is becoming a primary need (Noman, 2018). To cope with this need, the first solution arrived was to use Multiple Input Multiple Output (MIMO) technology to serve connecting devices simultaneously in (Marzetta, 2007). And the detailed model was presented in (Marzetta, 2010) and (Larsson, 2014). On the other hand, OFDM is a digital modulation technique that divides a high-speed data stream into multiple subcarriers, each carrying a low-speed signal. OFDM MIMO is a combined technique that uses multiple antennas at both the transmitter and receiver, along with OFDM's multiple subcarriers. OFDM MIMO technology did not just achieved high data rates but also ensured spectral efficiency. As the number of users in each year is increasing geometrically and also diversity in communication devices are being observed, more antennas and use of much high performing technology has become a crucial need. On that consequence, the concept of incorporating more transmitting and receiving antennas in MIMO came to serve more users at a time. But the main problem here was overhead in processing channel state information (CSI). We need more accuracy in CSI computation to eliminate interferences as well as reducing computational time. And that optimization is a NP hard problem indeed.Processing CSI refers as Channel Estimation. Channel estimation is very important as it informs the propagation effect of the signals which are propagated in MIMO systems through unobserved channel. The methods available in literature for channel estimation can be broadly divided into data-aided, blind and semi-blind methods. Blind and semi-blind methods do have iteration problem in random initialization (Antón-Haro, 1997). So, we are going to concentrate on data-aided methods only in this paper. The most popular data-aided channel estimation methods are Least Square (LS) and Minimum Mean Square Error (MMSE) estimation. The both methods lack in achieving required result when the number of antennas increase (Meiyan, 2017). Optimization of these methods using evolutionary algorithms has got popularity in these years. But this approach yet has research scope and that motivated us to find some better solution.

---

[1] Department of CSE, Gono Bishwabidyalay, Savar, Dhaka-1344, Bangladesh.
*Corresponding Author: email: shahriar.hassan303@gmail.com Phone:+8801778390534.

Our objective in this paper is to propose a robust and fast method for channel estimation in OFDM MIMO employing meta heuristic methods which will have reduced error rate. The rest of this paper is organized as follows: background and literature review are discussed in section 2. Details about the proposed method was presented in section 3. Section 4 contains the experiment setup and experimental result discussion. And last but not the least, we conclude in section 5.

## 2. Background and Literature Review

In practice, the communication path between two communicating nodes, i.e., Base Station (BS) and Terminal, in a wireless communication system is full of obstacles. So, all the transmitted signals may not follow a single path rather get reflected from many objects and get to the receiver. Thus, it forms a multipath propagation. The problem with the multipath propagation is the superposition of different versions of same signals or different signals at the receiver end. Further, superposition of signals can cause distortion in signals. Also, the signals can be refracted in many objects and thus the fading or attenuation is the result. Also, the channel may be noisy. These channel effect has an issue when receiver tries to demodulate the received modulated signals sent from senders. The resultant signal often becomes inaccurate with respect to original signal sent. If the receiver somehow knows the properties of the propagation channel in advance, then the issue can be solved to a satisfactory extant. Channel State Information (CSI) is the term used to define channel properties. And the method of gathering CSI is called Channel Estimation. The most common channel estimation technique is to transmit a known signal called Pilot Sequence from transmitting antennas and receiving antennas receive that signal and analysis the effect of multipath propagation, i.e. combined effect of scattering, fading, and power decay with distance. This technique is known as data aided channel estimation technique (Biguesh, 2006). The two dependable channel estimation techniques are described in the section 2.2 and 2.3 after the discussion on the data model.

## 2.1 Data Model

Let's assume there are $I$ cells in a cellular system. And there is only one BS at any cell and the BS contains an array of $P$ antennas. The number of users in a cell is $K$. The relation between $P$ and $K$ can be defined as $P \gg K$ (Marzetta, 2010). We assume that the $K_{th}$ terminal in every $i_{th}$ cell, where $i \in I$, synchronously transmits a pilot sequence $s_k$ comprising $\tau$ pilot symbols,

$$s_k = [s_{k1}, s_{k2}, s_{k3}, \ldots \ldots \ldots s_{k\tau}]^T \qquad (1)$$

All pilot sequences of all K users after combining presented as a matrix $S$ of size $\boldsymbol{K \times \tau}$.

$$S = [s_1, s_2, s_3, \ldots \ldots \ldots s_k] \qquad (2)$$

BS is assumed to have knowledge about all these mutually orthogonal pilot sequences. Because of the $K_{th}$ pilot sequences being transmitted, $i_{th}$ BS receives $P \times \boldsymbol{\tau}$ *signal*. That is,

$$Y = [y_1, y_2, y_3, \ldots \ldots \ldots y_P] = HS + N \qquad (3)$$

Here, the channel matrix of size $P \times K$ is represented with H and a noise vector of size K is represented with N in the formula (3). We assume the noise used here is thermal noise, i.e., white gaussian noise.

## 2.2 LS Estimator

LS estimator simply reduces the Mean Squared Error (MSE) between received pilot sequence and the estimated sequence. It can be formulated as:

$$H_{LS} = arg_{H_{LS}} \min ||Y - H_{LS}S||^2 \qquad (4)$$

The LS estimator's channel estimates for the channel impulse response are given by,

$$H_{LS} = YS^H(SS^H)^{-1} \qquad (5)$$

Here, the $S^H$ stands for S's conjugate transpose. It is possible to further simplify (5) by using the unitary property of conjugate transpose of a matrix as follows:

$$H_{LS} = YS^{-1} \qquad (6)$$

The problem with LS expressed in many literature is: though it is easy to implement and having lower computational complexity it does not provide approximate result that is very far from optimal one (Ghauri, 2013).

## 2.3 MSSE Estimator

An improved form of the LS channel estimation is the MMSE channel estimation. By determining a decent linear estimate in terms of the weighted vector and the value of the LS estimate, $H_{LS}$, the MMSE channel estimator reduces the MSE between the genuine channel, H. This can be mathematically written as:

$$H_{MMSE} = arg_{H_{MMSE}} \min ||H - H_{MMSE}||^2 \qquad (7)$$

Here, H denotes the reference channel matrix. As follows is the simplified MMSE estimate equation:

$$H_{MMSE} = WTT^{-1}Y \qquad (8)$$

Here, T is the autocorrelation, while W is the weight matrix.

MMSE is complex but gives better result as it employs some methods to reduce the effects of noise. But the computational complexity is much higher (Ghauri, 2013). As in wireless communication speed is much important so the computational overhead present in MMSE is unbearable.

## 2.4 Related Works

We have as far discussed about traditional methods of channel estimation LS and MMSE. And we also discussed the issues of LS and MMSE in previous sections. And so the recent concern is to optimize the methods, i.e. if we can optimize LS to a good extant without introducing much computational overhead than it is better to use optimized LS than more complex MMSE as MMSE is less speedy. For this reason, in many literature researchers has taken the help of meta-heuristic approach to optimize channel estimation. As far we have seen the approaches combined to existing algorithms used to optimize are genetic algorithms, particle swarm optimization (PSO), ant colony optimization and so on. In this section we have discussed few recent works on that direction.

Noman et. al. (Noman,2018) has proposed a method to optimize both LS and MMSE method using Genetic Algorithm (GA). They found that GA optimized LS method outperformed other methods with less complexity. But our observation is it can be further optimized by reducing number of iterations to convergence. In GA the local optimum is conserved in less number of iterations but can be far from global optimal.

Agarkar et. al. (Agarkar, 2018) proposed a method of channel estimation for OFDM MIMO employing Particle Swarm Optimization (PSO). That uses information from LS and Linear MMSE (LMMSE) to reduce error rate. It shown to be reduce Bit Error Rate (BER) at every Signal to Noise Ratio (SNR) level. Also, it converges fast.

Mandloi et. al. (Mandloi, 2015) proposed a congestion control ant colony optimization (CCACO) based symbol detection technique. Symbol detection is a technique that is highly dependent on CSI. It also reduced the BER.

Hei et. al. (Hei, 2013) proposed a combined method based on GA and PSO for symbol detection for MIMO. Vidhya et. al. (Vidhya, 2014) also proposed similar method to optimize channel estimation technique. But both of these methods use mutation for a large section of population having best fitness. Also, it reproduces the worst fitness samples randomly. Thus, it increases exploitation but lacks exploration of the solution space. Imbalance between exploration and exploitation makes convergence slower.

Kallur et. al. (Kallur, 2021) proposed a hybrid method based on PSO and Gravitational Search Algorithm (GSA). Only PSO treats channel estimation problem as a single objective problem. But they address the angle of signal propagation to make it multi-objective problem.

As we have discussed, in the existing works researchers used different evolutionary methods to optimize channel estimation. But, the balance between exploration and exploitation has never been concerned in these works. In our proposed model, we try to minimize the error rate by keeping balance between exploration and exploitation. Thus, our proposed work provides a novel direction in this research domain.

## 2.5 Problem Definition

Given a reference channel vector and initial estimation of channel state information (i.e., channel estimation using LS), we need to find a way to optimize the channel estimation such that:

    i. It will minimize the difference between estimated channel and reference channel

    ii. It will converge as early as possible

Here the measure for the difference between estimated channel and reference channel will be the Mean Squared Error (MSE) and convergence time will be measured with number of iterations in optimization process.

## 3. Proposed Method

The basic concept of GA is to take a set of individual solutions as a population and applying some analogous to genetics methods to find the best solution with respect to a fitness function. At each generation the worst solutions are discarded and recombination between better solutions progress the generation. This recombination technique is named as crossing over. Again, mutation is applied by slightly changing in solution vectors to add diversity and speed up the process of convergence. It thus ensures at-least a locally optimal result.

On the other hand, in PSO, a swarm of random particles (solutions) that resembles a flock of birds is initially placed in the search space. Each particle has a fitness value that is assessed by the objective function to be optimized, and it also has a velocity that determines the direction in which it will be "flying". The algorithm starts with a population and at each iteration it improves its position and also velocity according to three parameters: inertia, social and personal acceleration factor. It is popular to produce solution more close to global optima.

In our proposed solution we are trying to incorporate both the benefits from GA and PSO. At each generation we are going to do the following tasks:

a. Divide the population three sections: best, better and worst based on fitness.
b. Discard the worst solutions.
c. Mutate the best solutions.
d. Operate PSO on better solutions and reproduce the discarded solutions in next generation with the help of PSO knowledge.

The pictorial representation of the key steps of our proposed method is given below:

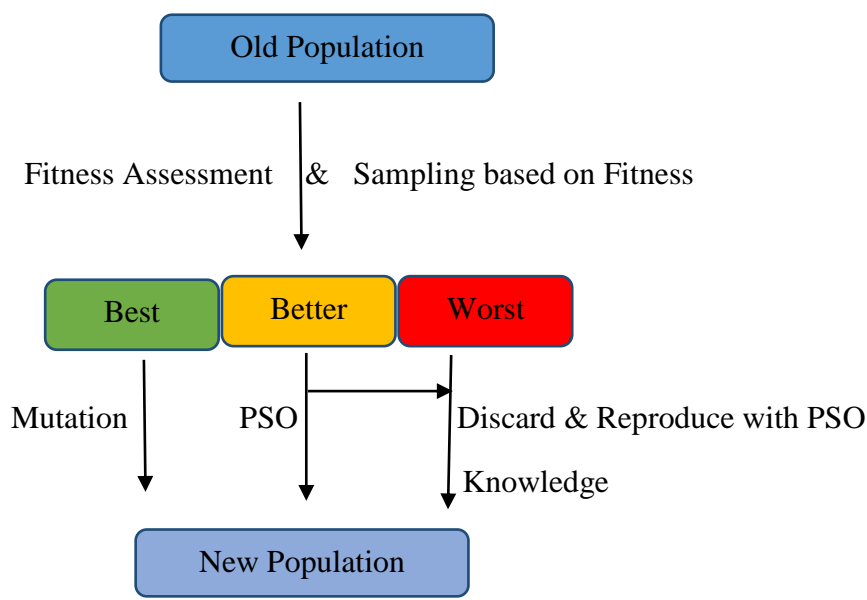

**Fig-1. Key Steps of Proposed Method.**

The purpose of mutating the fittest solutions is that they are more close to optima and slight change may accelerate their way to get the optimal result. The worst solutions are discarded as they are much far from optima and may add time to converge. But it introduces a risk of including much exploitation and decreasing exploration. So, the amount of mutation operation and reproduction should be bounded in affair limit so that it helps to balance between exploitation and exploration. In this manner we are getting more fit solutions at each iteration and thus reducing number of iterations. It is taking the advantages of both GA and PSO.

## 3.1 Problem to Concern

The following two points we have to concern to step towards the solution:

a. What will be the initial population?
b. What will be the fitness function?

Our goal is to optimize LS appropriately so that it will have less computational cost than MMSE and better or similar result as MMSE. So, the initial population will be the output of LS estimation. As we get the output of LS our fitness function will be:

$$Fitness = [(H - H_{LS})/H]^2 \qquad (9)$$

Our goal is to minimize the fitness function.

## 4. Experiment and Result Discussion

### 4.1 Experimental Setup

For the experimental purpose we have employed Monte Carlo simulation method; i.e., required data are generated randomly. We used Matlab R2018a software for simulation. For channel estimation we used the parameters in Table 1.

**Table 1: Channel estimation parameters.**

| Parameters | Specification |
|---|---|
| No. of Transmitting Antenna | 8 |
| No. of Receiving Antenna | 8 |
| Career | OFDM |
| Channel | AWGN, Slow Fading |
| Modulation | BPSK |
| Duplex Mode | TDD |

We have at first simulated LS and MMSE methods and then optimized LS with PSO. Then we have compared their performance with Signal-to-Noise Ratio (SNR) vs. MSE plotting. For PSO optimization we have used following parameters in Table 2.

**Table 2: Parameters used in PSO.**

| Parameters | Specification |
|---|---|
| Initial Population | 64 |
| Inertia (W) | Decreasing from 0.9 to 0.4 with iteration |
| Acceleration Factor (C1=C2) | 2 |
| Maximum No. of Iteration | 10 |
| Tolerance | $10^{-2}$ |
| No. of Runs | 12 |

In next step we have developed our proposed algorithm and compared with previous algorithms. For our proposed algorithm we have used following parameters in Table 3.

**Table 3: Parameters used in our proposed algorithm.**

| Parameters | Specification |
|---|---|
| Initial Population(Q) | 64 |
| Inertia and Acceleration Factor | Same as PSO |
| Mutated Individuals | 0.05Q |
| Reproduced Individuals | 0.05Q |
| Mutation | Gaussian Convolution with (0, 0.02) |
| Maximum No. of Iteration and Tolerance | Same as PSO |
| No. of Runs | 12 |

In our proposed method we have restricted introduction of mutation and reproduction to 0.1Q to eliminate worsen the performance introducing so much randomness. To reduce sorting cost, we have chosen the best individual at random with the fittest one and for the worst individuals also with the worst fit one with some random ones. Our reason to choose the variance as 0.02 is also validated through experiment and the results will be discussed in following section. The reason of choosing maximum iteration to 10 is that at maximum 10$^{th}$ iteration we get satisfactory results and we wanted to not adding much iteration cost.

## 4.2 Result Discussion

We developed our proposed algorithm and compared it with LS, MMSE estimator and the optimized LS estimator with PSO at different SNR values and we found the following result shown in Fig. 2.

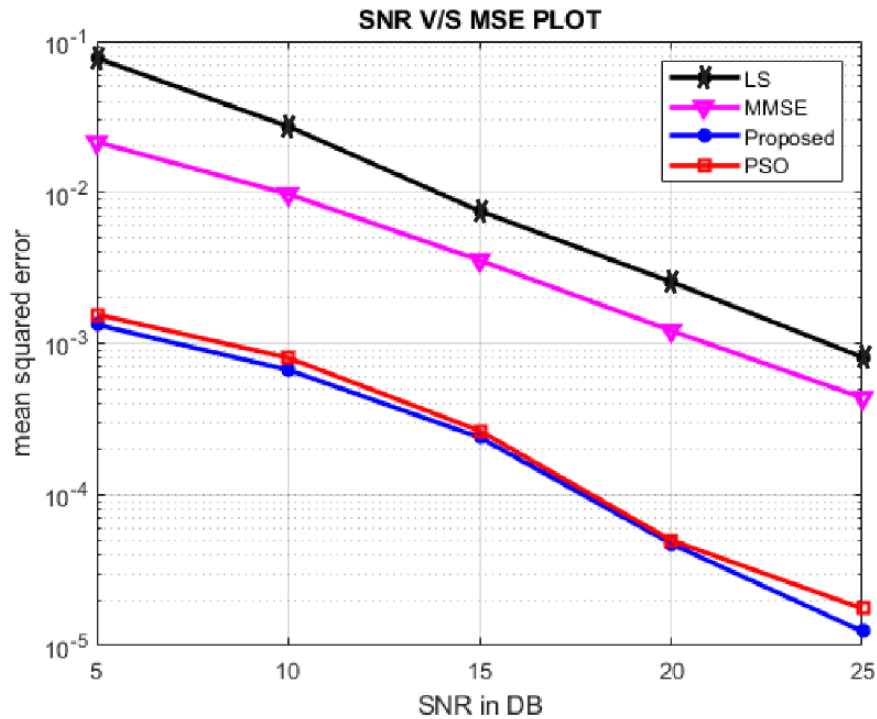

**Fig-2: SNR VS. MSE comparison among LS, MMSE, PSO and Proposed method.**

From Fig. 2, we can see that the proposed algorithm did better at some SNR values but we cannot definitely say that the proposed model outperforms PSO. Rather it is wise to say that the proposed model gives similar result as PSO on our measurement verdict MSE. We also can see that both PSO and proposed optimization has outperformed the complex state of the art algorithm MMSE and hence our optimization is successful. That was the satisfaction to our first objective.

Now if we have similar result as PSO then why to use proposed algorithm, a big question. Our experiment shown that the proposed method converges at least iteration compared to PSO. The result of our experiment is shown in Table 4.

**Table 4: Number of iterations needed for convergence at different SNR Level with PSO vs. our proposed method.**

|                | PSO     |         | Proposed Method |         |
|----------------|---------|---------|-----------------|---------|
| SNR Level (db) | Maximum | Minimum | Maximum         | Minimum |
| 5              | 8       | 0       | 4               | 0       |
| 10             | 8       | 0       | 8               | 0       |
| 15             | 10      | 4       | 9               | 0       |
| 20             | 10      | 0       | 10              | 1       |
| 25             | 10      | 2       | 10              | 3       |

From the above table we can see that though our proposed method converges later than PSO in the case of higher SNR, it converges earlier or at the same time as PSO for lower SNR level. As we predicted that because of the modification we did on PSO it will move towards the optimal result faster hence proven with this.

To understand the exploitation and exploration nature of PSO and our proposed method we plotted the values of channel vector at each iteration and the result is shown in Fig. 6. From the figure we found that our proposed method is more explorative than PSO because of reproduction of worst solution in following generation. Also, it is more exploitative on the optimal region. This has a significance that it tries to find more optimal result when close to optima. It also converges to optima earlier because of discarding worst solutions and restarting from their personal best and boosting the fittest solutions through mutation.

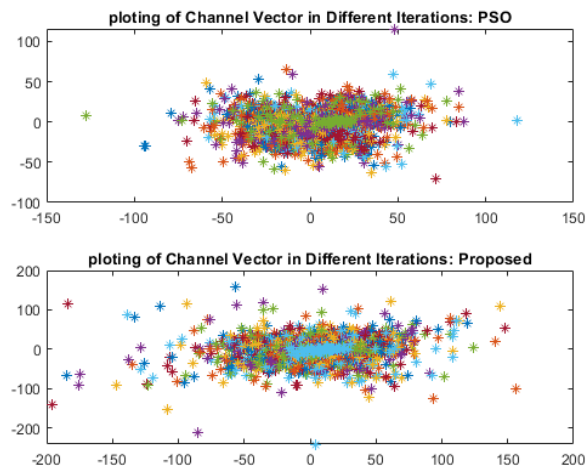

**Fig.-3: Exploitation vs. Exploration nature of PSO and Proposed Method.**

## Conclusion and Future Work

In wireless communication channel estimation is very crucial. Computation time is also a very important factor especially in the case of TDD. Correctness of symbol detection and encoding very much rely on CSI. As our proposed method minimize the error in channel estimation to a great extant with minimal computational complexity we hope that it will contribute to the sector. Getting optimal result with lesser iteration is the motivation of our future work.